\documentclass[aps,prl,twocolumn,superscriptaddress,showpacs,floatfix]{revtex4}

\usepackage{graphicx}

\begin{document}

\title{First-principles Studies for the Hydrogen Doping Effects on Iron-based Superconductors}

\author{Hiroki Nakamura}
\email[]{nakamura.hiroki@jaea.go.jp}
\author{Masahiko Machida}
\email[]{machida.masahiko@jaea.go.jp}
\affiliation{CCSE, Japan Atomic Energy Agency, , 6--9--3 Higashi-Ueno,
Taito-ku Tokyo 110--0015, Japan}
\affiliation{CREST (JST), 4--1--8 Honcho, Kawaguchi, Saitama 332--0012,
Japan}
\affiliation{JST, Transformative Research-Project on Iron Pnictides (TRIP), Chiyoda, Tokyo 102-0075, Japan}

\date{\today}

\begin{abstract}

We study hydrogen doping effects in an iron-based superconductor LaFeAsO$_{1-y}$ by using the 
first-principles calculation and explore the reason why 
the superconducting transition temperature is remarkably enhanced by the hydrogen doping. 
The present calculations reveal that a hydrogen cation stably locating close to an iron atom 
attracts a negatively-charged FeAs layer and results in structural distortion 
favorable for further high temperature transition. 
In fact, the lattice constant $a$ averaged over the employed supercell
shrinks and then the averaged As-Fe-As angle $\alpha$ approaches 109.74$^\circ$ with 
increasing the hydrogen doping amount.  Moreover, the calculations clarify electron doping effects 
of the solute hydrogen and resultant Fermi-level shift. 
These insights are useful for design of high transition-temperature iron-based superconductors.

\end{abstract}

\maketitle

Since the discovery of the high-temperature superconductivity in LaFeAsO$_{1-x}$F$_x$ \cite{kamihara},
iron-based superconductors have drawn wide attention due to 
not only their high-temperature transition but also peculiar superconducting properties.
According to the accumulated insights, most of iron-based superconductors require 
chemical doping like substitution as LaFeAsO$_{1-x}$F$_x$
or deficiency as LaFeAsO$_{1-y}$ for their optimal high-$T_{\rm c}$ superconductivity \cite{ishida}.
In addition, most of their mother compounds exhibit the 
stripe-type antiferromagnetic ordering just after the orthorhombic transition 
on cooling the temperature. There are several similarities with cuprate high-$T_{\rm c}$ superconductors 
but some original particularities. Therefore, more systematic research is 
now still in great demand. 

Recently, Miyazawa et al.,\cite{miyazawa} reported a fascinating fact 
that the transition temperature $T_{\rm c}$ of LaFeAsO$_{1-y}$
is enhanced from 28 K to 38 K by doping the hydrogen in its
high-pressure synthesis. It surprisingly corresponds to more than 30\% of $T_{\rm c}$ enhancement, which is 
unlikely in other superconductors.  
This new synthesis scheme is quite different 
from the previous chemical doping
such as substitutions and deficiencies, because the enhanced $T_{\rm c}$ 
is much beyond the optimum reach of so-called La-1111 compounds. 
Therefore, the hydrogen doping effect should be an important key to 
the superconducting mechanism
and $T_{\rm c}$ enhancement in iron-based superconductors. 
In this paper, we study the hydrogen doping effects through systematic first-principles calculations.
The highlights of this paper are to clarify at which the doped hydrogen atom 
prefers to stay inside the crystal unit 
and how the doped hydrogen atoms deform the crystalline structure and resultant electronic one.
Most of the calculated results are consistent with measurement data, while 
they contain novel insights never experimentally accessible. 
 
Before starting the calculation study of the hydrogen doping effects, there are a few questions 
to narrow the point. 
The first one is whether the hydrogen atoms really keep 
their positions inside the crystal.
The nuclear magnetic resonance measurement on the 
hydrogen (H-NMR) confirmed the existence of doped hydrogen atoms \cite{miyazawa,mukuda} not 
at any surfaces and boundaries but inside the crystal. 
This indicates that there are stable locations inside the crystal for the hydrogen atom. 
The next one is 
internal locations of the doped hydrogen atoms.
One naively expects that doped H atoms fills O vacancies.
Contrary to this conjecture,
Fe-NMR verified that the hydrogen atoms partly locate 
close to Fe atoms \cite{mukuda}.
However, their exact locations have not been known yet because of 
its technical difficulties.
Then, we explore the exact position at which the hydrogen atom stably
locates around Fe atom by first-principles calculations.
Once one solves the above problem, the final goal is 
to clarify how the hydrogen atom raises $T_{\rm c}$.
In this paper, we reveal some important roles of the doped 
hydrogen atoms. The first one is that 
the hydrogen atoms tend to shrink lattice constants 
like the replacement of Ln element in LnO blocking layer of 1111 compounds.
The superconducting transition temperature is known to be quite sensitive 
to the variation of the lattice constants.
Moreover, it is well-known that the As-Fe-As angle, $\alpha$ 
and the height of an As atom from an Fe-plane, $h_{As}$
are closely correlated with $T_{\rm c}$ \cite{lee,takano}. 
In fact, we find that the doped hydrogen distorts 
the crystal structure more suitable for high-$T_{\rm c}$ superconductivity. 
The second one is whether the hydrogen doping
supplies additional carriers. 
The calculations reveal that the hydrogen 
provides electron and the Fermi surfaces change
their shapes in a large doping level. 

\begin{figure}
\includegraphics[width=8.3cm]{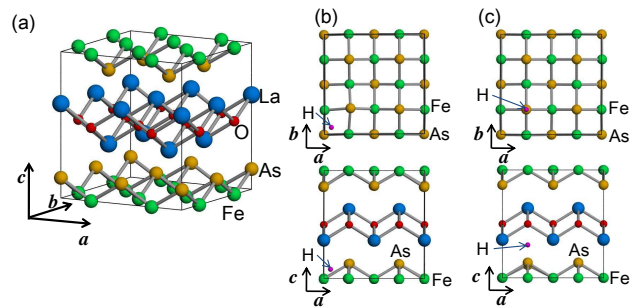}
\caption{
(a) $2\times 2\times 1$ supercell of LaFeAsO. (b) Most stable structure of (LaFeAsO)$_8$H, in which
a H atom exists in the FeAs layer. (c) Metastable structure of  (LaFeAsO)$_8$H which contains a H atom
around the LaO layer.}
\label{fig1}
\end{figure}

Let us briefly show the present calculation techniques. 
We examine hydrogen-doping effects in iron-based superconductors based on 
the density functional theory (DFT).
The calculation package employed throughout 
this paper is VASP \cite{vasp}, which adopts PAW method\cite{paw} 
and GGA \cite{pbe} exchange-correlation energy.
In order to examine how the doped hydrogen deforms the crystal structure, we prepare 
$2\times 2\times 1$ supercell of (LaFeAsO)$_8$ as shown 
in Fig. \ref{fig1}(a).
If one puts a hydrogen atom into the supercell, the chemical formula is given by LaFeAsOH$_x$, in which $x=0.125$.
Accordingly, $x = 0.25$ when inserting two H atoms.
Thus, we can control the doping level $x$, although the variation range is quantized. 
In order to find relaxed optimal positions for the hydrogen atoms inside the supercell,
we optimize the crystal structure of LaFeAsOH$_x$, so that
the forces on all atoms become less than 0.01 eV/\AA.
The parameters set in the DFT calculations are given as follows.
$k$-points are taken $4\times 4 \times 4$, and
the self-consistent loops are repeated 
until the energy deviation becomes less than $10^{-5}$ eV with the cut-off energy being 500 eV.

Now, let us present the calculation results.
First, we explore stable positions for a hydrogen atom inside LaFeAsO crystal unit.
We initially prepare a few tens configurations, in which an H atom is randomly distributed 
inside the LaFeAsO unit cell, and optimize their structures.
Then, the doped hydrogen atom finds out a stable location.  
As a result, we find two typical locations for an additive H atom. 
The first type is around an alternately vacant As position which 
is mirror symmetric with a non-vacant As across the Fe-plane. 
According to the calculation result, the exact stable location slightly shifts from 
the vacant As position as shown in Fig. \ref{fig1}(b). 
The second one is just below a La atom alternately symmetric across the O plane 
as shown in Fig. \ref{fig1}(c).
The former case is found to be more stable by comparing their total energies.
This result actually agrees with Fe-NMR measurements \cite{mukuda}.
However, the calculated energy difference is 50 meV per LaFeAsOH$_{0.125}$
and 
the hydrogen atom may partially occupy the second position as 
a quasi-stable location.

Here, we turn to O vacancies.
As mentioned above, H atoms are expected to enter O vacancies.
In order to examine this,
we compare the total energies of the following two types of (LaFeAs)$_8$O$_7$H structures.
A configuration is that a H atom fills the O vacancy, and the other one is that 
it locates near an Fe atom without filling the O vacancy.
As a result, more stable configuration is found to be the former one.
Therefore, we conclude that H atoms first tend to fill O vacancies.
In this case, the H doping corresponds to just a substitution like LaFeAsO$_{1-x}$F$_x$,
and it is found to be not relevant to more than 30 \%
$T_{\rm c}$ enhancement.
In fact, the Fe-NMR result \cite{mukuda} insists that
H atoms exist near Fe atoms.
This indicates that an excess portion of the doped H atoms located in FeAs layers.
Hereafter, we focus on the H locations near Fe atoms as a new mechanism of
$T_{\rm c}$ enhancement.

\begin{figure}
\includegraphics[width=8.3cm]{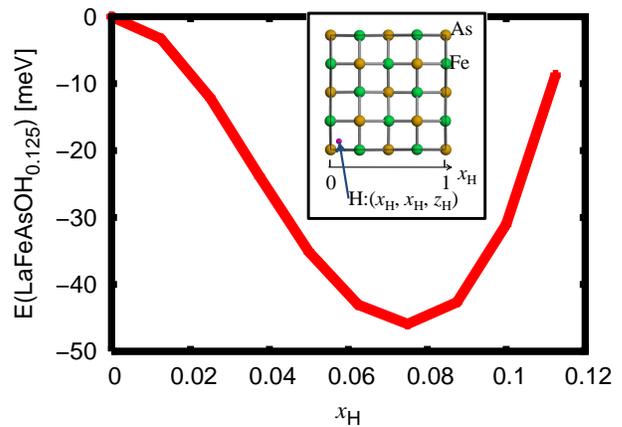}
\caption{
Total energy per LaFeAsOH$_0.125$ as a function of inner atomic coordinates of the H atom in (LaFeAsO)$_8$.
Inset shows the position of the H atom. Inner atomic coordinate $z_{\rm H}$, which is along $c$-axis, is optimized.
}
\label{fig2}
\end{figure}

Let us concentrate on the location inside FeAs layers.
As shown in Fig. \ref{fig1}(b), 
we find out that the most stable position shifts to (0.07,0.07) in the inner atomic coordinate 
from (0,0) corresponding to just above an Fe atom. 
We carefully trace a variation of the total energy with respect to the position shift 
as $(x_{\rm H},x_{\rm H},z_{\rm H})$, in which once $x_{\rm H}$ is fixed $z_{\rm H}$ is optimized.
Figure \ref{fig2} shows the total energy variation as a function of $x_{\rm H}$. 
At $x_{\rm H}=0$, the total energy counterintuitively shows maximum.
The present calculation reveals the reason as follows. 
The position $x_{\rm H}=0$ is actually unstable 
due to the repulsion from the nearest La$^{3+}$ cation for the H$^{+}$ one.
Then, it shifts to avoid the so-called Coulomb-repulsion pressure
from the La$^{3+}$ cation.
This result obviously indicates that Ln cation in 1111 crystals 
has an important influence on FeAs layer.
The Ln cation and FeAs layer keep exquisite balance.
On the other hand, we notice in Fig. \ref{fig2} that 
the energy difference in the energy valley is not 
so large that the hydrogen atom can largely vibrate around the 
stable position. Such a localized dynamical effect 
is beyond the present study.

\begin{figure}
\includegraphics[width=8.3cm]{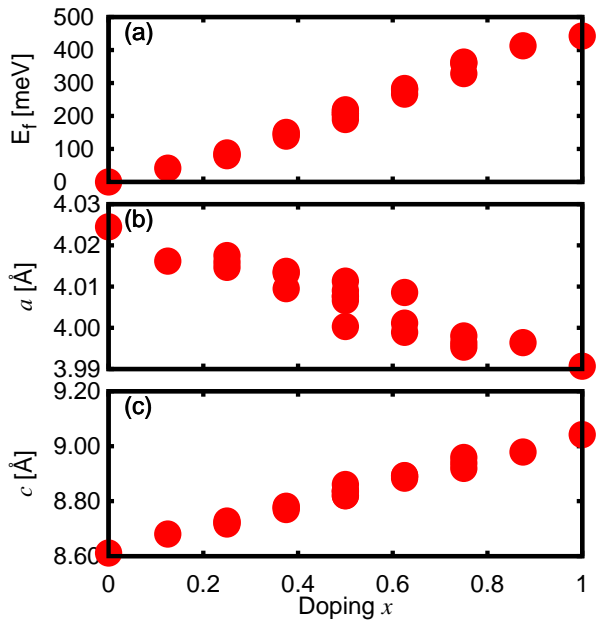}
\caption{$E_{\rm f}(x)$ (see Eq. (\ref{eq1})) (a),
lattice constants $a$ (b) and $c$ (c) as a function of doping $x$.}
\label{fig3}
\end{figure}

Next, we examine the hydrogen-doping amount dependence.
Since we now know the most stable location inside the crystal unit, as initial configurations we put 
H atoms on the positions as shown in Fig.\ref{fig1}(b). 
There are 1 to 8 As atoms inside the present super-cell.
This indicates that there are also 8 possible initial locations for an H atom.
Then, the number of possible configurations are totally $2^8=256$.
Among them, several configurations are symmetrically equivalent to others, and 
the initial sets can be then reduced to 22 patterns, among which 
the number of doped hydrogen atoms is ranged from 1 to 8.
We calculate stable structures and total energies on these all 22 patterns.
Figure \ref{fig3}(a) shows an energy $E_{\rm f}(x)$ as a function 
of the doping level $x$, which is obtained by dividing the number of H atoms in 
the cell by 8. The energy $E_{\rm f}(x)$ is defined as
\begin{equation}
E_{\rm f}(x) = E({\rm LaFeAsOH}_x)-E({\rm LaFeAsO})-\frac{x}{2}E({\rm H}_2),
\label{eq1}
\end{equation}
where $E(\cdots$) stands for the total energy of the correspondent chemical formula.
As seen in Fig. \ref{fig3}(a), $E_{\rm f}(x)$ is positive and increases with $x$.
One finds that all the present possible doping 
is energetically unstable compared to non-doped LaFeAsO.
This is related to a fact that high pressure synthesis is needed for doping.
The doping $x$ variation of the lattice constants $a$ and $c$ are shown 
in Figs. \ref{fig3}(b) and (c), respectively.
From both the figures, one finds that $a$ decreases with increasing $x$.
The observed $a$ of H-doped LaFeAsO$_{1-y}$ is 3.9938 \AA \cite{miyazawa}, 
which is distinctly smaller than that of the non-doped LaFeAsO$_{1-y}$ (4.0257 \AA)\cite{miyazawa2}.
This fact is consistent with our calculations.
On the other hand, the observed $c$ counterintuitively decreases from 8.7190 \AA \cite{miyazawa2} to 8.6898 \AA \cite{miyazawa}
by H doping, although the lattice constant $c$ increases with $x$ in the present calculation. 
The present tendency contradicts the measurement data.
We suggest a reason on the opposite tendency as follows.
It is well-known that the hydrogen can easily diffuse 
in various compounds because the energy barrier for the diffusion 
is not so high as seen in Fig.\ref{fig2}. This means that any hydrogens do not always occupy the 
energy minimum point. For instance, if a hydrogen atom partially occupies the second energy minimum location as 
shown in Fig.\ref{fig1} (c), then $c$-axis constant may rather decreases, because the neighbor 
is negatively-charged O$^{2-}$.
Although such statistical analyses on various configurations including the 
diffusion effects are beyond the present work, the $c$-axis shrinkage clearly indicates 
that doped hydrogen atoms partly stay at the positions close to the anions. 

\begin{figure}
\includegraphics[width=8.3cm]{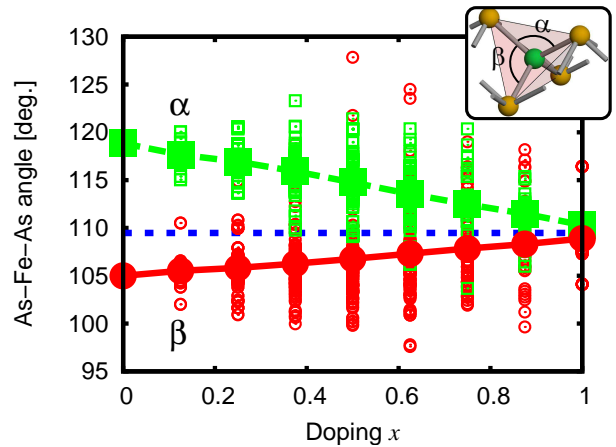}
\caption{As-Fe-As angles as a function of doping $x$.
Small squares and circles stand for $\alpha$ and $\beta$ angles, respectively,
 which are depicted in the inset.
Large squares and circles are the average of $\alpha$ and $\beta$, respectively.
Horizontal line denotes the regular tetrahedron angle, 109.47$^\circ.$
}
\label{fig4}
\end{figure}

Figure \ref{fig4} is the highlight of the present paper.
Lee et al., \cite{lee} pointed out that $T_{\rm c}$ is closely correlated 
to the As-Fe-As angle, $\alpha$ (see the inset of Fig. \ref{fig4}, where $\beta$ is noted to be also
another equivalent index).
$T_{\rm c}$ takes the maximum value when $\alpha(\beta)=109.47^\circ$, at which FeAs$_4$ tetrahedron becomes 
a perfectly regular one.
As seen in Fig. \ref{fig4},
the averaged $\alpha(\beta)$ clearly approaches to the optimal value with increasing 
the hydrogen doping ratio $x$.
The reason is given as follows.
The hydrogen ion (cation) attracts negatively-charged FeAs layer, and 
then the in-plane lattice constants decrease.
Consequently, the As height increases and the angle $\alpha(\beta)$ decreases(increases).

\begin{figure}
\includegraphics[width=8.3cm]{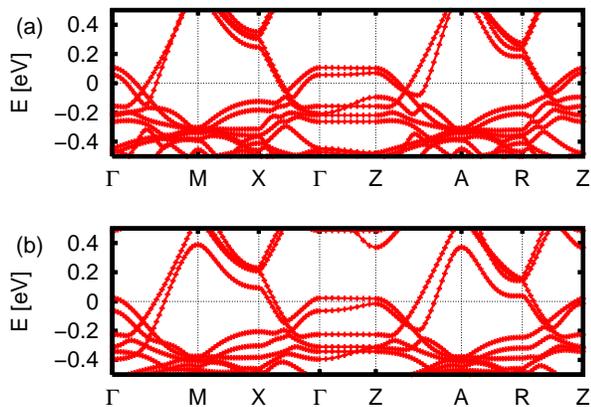}
\caption{
Band structure of LaFeAsOH$_x$ of (a) $x=0.125$ (b) $x=0.25$.
The Fermi energy is set to zero.
}
\label{fig5}
\end{figure}

The next issue is an effect of the hydrogen doping on the electronic band structure.
We display band dispersions of H-doped LaFeAsO in Fig. \ref{fig5}.
It is noted that the current Brillouin zone (BZ) is folded 
to a quarter due to the use of $2\times 2$ supercell.
In the conventional unit cell, the iron-based superconductors generally
have hole pockets around the $\Gamma$ point (the center of BZ)
and electron pockets around $M$ point (the corner of BZ). 
In the present zone, $M$ point moves to $\Gamma$ one by the zone folding, and both the pockets almost 
overlap around $\Gamma$ one.
Actually, both the pockets crosses in the case at $x=0.125$ as displayed in Fig. \ref{fig5}(a), i.e., the Fermi surfaces 
between the hole and electron pockets are well nested.
Meanwhile, the hole pocket almost disappears at $x=0.25$ as shown in Fig. \ref{fig5}(b).
The reason is that the doped hydrogen atom becomes a cation and supplies an electron.
Consequently, the Fermi level lifts up and the hole pocket disappears beneath the Fermi level.
This indicates that too much hydrogen doping drastically changes the electronic structure around the Fermi level and eventually diminishes the high-temperature superconductivity.

Finally, we discuss the hydrogen doping effects together with 
various knowledge on hydrogen additives in solid states.
First, we stress that the doped hydrogen shrinks all the lattice constants although 
extra element additives normally expand lattice constants. 
One of the origin is that the hydrogen cation locating stably close to Fe atom
attracts negatively-charged FeAs layer.
Thus, the mother compound LaFeAsO crystal is found to be rather loosely packed 
in contrast to very low solubility of the hydrogen in pure Fe crystal.
This fact is deeply relevant to the large pressure and substitution effects in 1111 compounds. 
From this discussion, it is expected that the hydrogen doping effect
on Sm-, Gd- and Nd-1111 compounds leads to further $T_{\rm c}$ enhancement.
Unfortunately, such enhancement was not reported, e.g., in H-doped SmFeAsO$_{1-y}$ \cite{shirage,hanna}.
However, in their compounds, it is noteworthy that the hydrogen is also soluble,
because their FeAs structures and associated electronic ones are still controllable.
Second, we point out that the doped hydrogen supplies an electron.
Namely, the hydrogen doping lifts up the Fermi level and 
the excess doping loses the nesting situation which is regarded as an electronic-structure 
stage for high-$T_{\rm c}$ superconductivity.
This means that there is the best hydrogen doping ratio while further $T_{\rm c}$ enhancement 
can be expected if one controls the carrier number by different ways from the 
hydrogen doping,
because the angle $\alpha$ monotonically approaches the best angle with the doping amount.
Third, we would like to mention that the local distortion generated by the doped hydrogen 
considerably varies inside the supercell unit and just the averaged distortion
is preferable for the transition temperature enhancement as shown in Fig. \ref{fig4}.
Accordingly, the emergent superconductivity seems to reflect 
not local atomic-scale but widely averaged distortions and resultant electronic-structure changes.
Since high diffusivity of the hydrogen is largely suppressed 
in a low temperature range as the superconducting transition, it is a big mystery that the interstitial hydrogen close to Fe seems not to work as a pair breaking potential. 

We have investigated the optimized crystal structure of the hydrogen doped
LaFeAsOH$_x$ by using the density functional theory.
The calculation results indicate that 
the hydrogen atom prefers to stay close to Fe atom.
In the case, the supercell calculations revealed 
that the averaged lattice 
constant $a$ shrinks with increasing the doping amount, which is consistent with
the experimental results. 
Meanwhile, the lattice constant $c$ 
expands within the present calculations, although 
it also shrinks in experiments. 
Therefore, it is considered that 
the hydrogen atoms locates at not only the most stable point 
but also other minimum ones.
The averaged structural distortions coincide with 
the correlation tendency of $T_{\rm c}$ on the crystal structure.
As a typical example, we observed that the As-Fe-As angle 
$\alpha$ approaches closer to 109.47$^\circ$ by increasing the amount of the 
doped hydrogen. However, the doping shifts up the Fermi level, because  
the hydrogen also has the carrier doping effect.
In conclusion, the present first-principles studies clarified some fundamental aspects 
in the hydrogen doping
beyond the findings obtained by experiments.
We believe that their insights are essential to materials design for further high-$T_{\rm c}$ superconductivity.

\end{document}